\documentclass{article}
\usepackage{graphicx}

\title
  {Seasonal and Lunar Month Periods Observed in Natural Neutron Flux at High Altitude}

\author{
YURI STENKIN,$^{1,2}$ VICTOR ALEKSEENKO,$^1$ ZEYU CAI,$^3$ \and ZHEN CAO,$^4$ CLAUDIO CATTANEO,$^5$ SHUWANG CUI,$^3$ \and ELIO GIROLETTI,$^{5,6}$ DMITRY GROMUSHKIN,$^2$ XUEWEN GUO,$^3$ \and CONG GUO,$^{4,7}$ HUIHAI HE,$^4$ YE LIU,$^8$ XINHUA MA,$^4$\thanks{Corresponding author: Xinhua Ma, Key Laboratory of Particle Astrophysics, Institute of High energy Physics, Chinese Academy of Sciences, P.O.Box 918(3), Yuquan Road 19(B), Beijing 100049, China E-mail: maxh@ihep.ac.cn } \and OLEG SHCHEGOLEV,$^1$ PIERO VALLANIA,$^{9,10}$ \and CARLO VIGORITO,$^{10,11}$ and JING ZHAO$^4$ \\
  $^1$ Institute for nuclear Research, Russian Academy of Sciences, \and Moscow, Russia \and
  $^2$ National Research Nuclear University MEPhI (Moscow \and Engineering Physics Institute), Moscow, Russia \and
$^3$ The College of Physics Science and Information Engineering, \and Hebei Normal University, Shijiazhuang, China \and
$^4$ Key Laboratory of Particle Astrophysics, Institute of High energy  \and Physics, Chinese Academy of Sciences, Beijing, China \and
$^5$ Dip. Fisica, Universit a Studi di Pavia, Pavia, Italy \and
$^6$ Istituto Nazionale di Fisica Nucleare, Pavia, Italy \and
$^7$ University of Chinese Academy of Science, Beijing, China \and
$^8$ The School of Physics, Shandong University, Jinan, China \and
$^9$ Osservatorio Astrofisico di Torino dell'Istituto Nazionale di \and Astrofisica, Torino, Italy \and
$^{10}$ Istituto Nazionale di Fisica Nucleare, Torino, Italy \and
$^{11}$ Dipartimento di Fisica dell'Universit\`a di Torino, Torino, Italy
 }

\begin{document}

\label{firstpage}

\maketitle

\begin{abstract}

 Air radon concentration measurement is useful for research on geophysical effects, but it is strongly sensitive to site geology and many geophysical and microclimatic processes such as wind, ventilation, air humidity and so on that induce very big fluctuations on the concentration of radon in air. On the contrary, monitoring the radon concentration in soil by measuring the thermal neutron flux reduces environmental effects. In this paper we report some experimental results on the natural thermal neutron flux as well as the concentration of air radon and its variations at 4300 m a.s.l. These results were obtained with unshielded thermal neutron scintillation detectors (en-detectors) and radon monitors located inside the ARGO-YBJ experimental hall. The correlation of these variations with the lunar month and 1-year period is undoubtedly confirmed. A method for earthquakes prediction provided by a global net of the en-detectors is currently under study.
\end{abstract}

keywords: thermal neutron; radon; lunar month periods; seasonal periods

\section{Introduction}

It is well known that radioactive gas $^{222}$Rn originated from the radioactive chain of $^{238}$U in soil and construction materials as inert gas can penetrate indoor environments and accumulate in basements and underground Laboratories. Its concentration in air can be rather high and it is strongly sensitive to site geology and many geophysical and microclimatic processes such as local seismic activity, air pressure and wind, tides and so on. A lot of work has been done to measure and keep under control the environmental radon concentration. Unfortunately, air radon concentration measurements are hampered by very big fluctuations due to air movement, ventilation, air humidity, etc. On the other hand, radon and its daughter nuclides are $\alpha$-active and therefore there is a neutron flux in equilibrium with radon in the Earth's crust due to nuclear ($\alpha$ ,n)-reactions on light nuclei such as Be, F, B, Na, Mg, Al, Si, etc. (not in air where no suitable targets for ($\alpha$ ,n)-reactions exist). These reactions produce fast neutrons with energy of a few MeV and are afterwards moderated down to thermal and epithermal energies covering several meters underground. The radon diffusion length in soil strongly depends on the site geology, ground water level, etc., a reasonable expectation being of the order of tens of meters. This feature of radon (and partially thoron) makes produced neutron flux sensitive to some geophysical phenomena such as local seismic activity, tides, etc. Any change of radon diffusion, including soil water level change, will cause change in neutron generation. Most of neutrons are captured by surrounding nuclei, but some of them can escape from absorption and come to air. Therefore, it is possible to monitor the radon concentration in soil by measuring the thermal neutron flux. This method developed by us (ALEKSEENKO et al.,2009; ALEKSEENKO et al. 2010) has many advantages, mainly the fact that it is not sensitive to air drought, ventilation, humidity, etc., but only to the radon concentration underground (in soil, rock, concrete, etc.). More precisely, the neutron yield is proportional to the radon density flux through soil, rock, concrete or other porous materials close to the detector. On the other hand, the higher radon density flux through soil, the higher should be radon concentration in air. This is why some correlation should exist between data obtained with radon meter and with en-detector located at the same site. In this paper, we report some results obtained by applying this method to our detectors running at high altitude.

\section{Experimental Setup}

Data have been collected since January 2013 in Yangbajing (Tibet, China) at an altitude of 4300 m a.s.l. inside the ARGO-YBJ experimental hall (D'ETTORRE BENEDETTO 2011). The results presented here have been obtained after late August, 2013 when the detectors configuration was changed. The natural thermal neutron flux variations have been measured using the small detector array PRISMA-YBJ (BARTOLI et al. 2016)) made by 4 en-detectors. The so-called en-detectors, developed in Institute for nuclear Research (INR), Russian Academy of Sciences (RAS) initially for cosmic ray study, are capable of measuring thermal neutrons and multiple passage of charged particles (STENKIN, 2010). The en-detector is sensitive to thermal and epithermal neutrons by means of a thin layer of inorganic scintillator ZnS(Ag) alloyed with $^6$LiF. $^6$Li has a high cross section for the (n, $\alpha$)-reaction: $^{6}Li + n \rightarrow ~^{3}H + \alpha + 4 .78 MeV$.  Heavy charged particles $^3$H and $\alpha$  produce point-like ionization with the emission of $\sim$ 160000 photons in the ZnS(Ag) scintillator. Light is collected by a 5-inchs photomultiplier tube (PMT) FEU-200 obtaining a signal of about 100 photoelectrons from the PMT's photo cathode. Significant features of the scintillator are existence of several time constants (from 40 ns to hours) and sensitivity to particle velocity: slowly moving heavy particles such as $\alpha$ or $^3$H excite more efficiently slow components in the detector pulse shape. The scintillator, with a surface of 0.35 $m^2$, and the PMT are located inside a black plastic tank of 200 l volume used as detector housing. Data acquisition includes a full pulse shape digitizer with a flash analog-to-digital converter (FADC) installed in a PCI slot of the On-Line PC. The data of each detector are collected and stored every 5 minutes.

Due to the very thin scintillator layer (only 30 $mg/cm^2$), charged particles (electrons, muons, etc.) lose only $\sim$60 keV and their signals are below the FADC threshold set at 150 keV. Only neutron captures and synchronous passages of several charged particles can be detected, and this results in a very low counting rate.  Two trigger systems of the array are implemented: 1) the coincidence of any 2 out of 4 detectors for EASs and 2) any hit of any detector with a conversion factor reduced by a factor of 16 to decrease the dead time. For variations study we use the second one with a counting rate of $\sim$ 2 Hz.

The efficiency of our scintillator for thermal neutron detection was found experimentally by neutron absorption. We measured the counting rate of our scintillator layer, then we put a similar layer under the first one (with a black paper between them) as an absorber and measured the counting rate again. Finally we compared the results: the obtained scintillator efficiency is $\sim$ 20 \%. A similar efficiency was also obtained by means of a simple Monte-Carlo simulation based on GEANT4 code. Simple analytical calculation using known neutron capture cross sections and scintillator thickness gave a similar result. Using the detector pulse shape analysis (see details in (ALEKSEENKO et al. 2015)), we can identify three signal types: "neutrons" produced mostly by thermal neutron captures by target nuclei $^6$Li, "charged" caused by multiple charged relativistic particles passage and "very slow" due to electromagnetic noise. The signal separation is possible due to different charge collection times in case of passage of charged relativistic particles and of slowly moving $\alpha$ and $^3$H emitted after the neutron capture by $^6$Li nuclei. All these signal types are counted and stored separately.

The sensitivity of the en-detector to radon concentration in air is produced by cascades of $\gamma$s during $\beta$-decays of radon daughter nuclei, mostly $^{214}$Bi and $^{214}$Pb. The sensitivity of en-detectors to thermal neutrons was checked experimentally using a $^{252}$Cf neutron source with a polyethylene moderator. No difference was found in the "neutron" channel between the energy deposited spectra with and without the source of thermalized neutrons, while the neutron counting rate with $^{252}$Cf was found to be coherent with the source activity. The sensitivity of the en-detector to radon ($^{220}$Rn or thoron) concentration in air through the "charged" channel was checked by means of a thoron source ($^{232}$Th).  In our ¡°neutrons¡± channel measurements we can estimate the fraction of radon originated neutrons vs that of cosmic rays by using barometric coefficient. Assuming that it is equal to 1 \% /mm Hg for cosmic ray branch and it is equal to 0 (in first approximation) for radon. This fraction is close to 10 - 20 \% for surface detectors. If so, the observed wave amplitudes being recalculated to soil radon variation amplitudes are expected to be higher by a factor of 5 - 10.

\begin{figure}
  \includegraphics[width=1\textwidth]{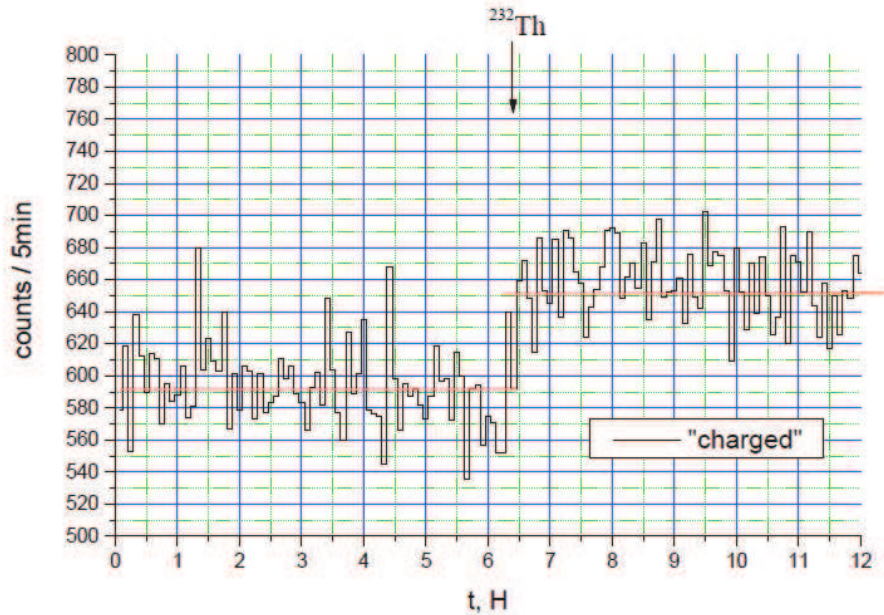}
  \caption{Counting rate of the "charged" channel of the en-detector as a function of time. The effect of the $^{232}$Th source placed near the detector after 6.4 hours is clearly visible.}
  \label{fig1}
 \end{figure}

Fig. \ref{fig1} shows the 5-minutes time series of the counting rate in "charged" channel, which indicates that the "charged" channel was found to be sensitive to radon (thoron) concentration in air. It should be noted that in a case of thoron gas the sensitivity of the detector to its concentration is provided by its daughters $^{212}$Pb and $^{212}$Bi $\beta$-decays similar as for radon-222. Sure we cannot distinguish between neutrons produced through different $\alpha$-emitters. We measure an integral flux produced in all nuclear reactions. Below we'll speak about a correlation between measured neutron and radon concentrations because only radon (both Rn-222 and Rn-220) as inert gas can penetrate through porous media (producing neutrons there directly or through its daughters decays) along with other soil gases to atmosphere thus giving rise to air radon measuring by a radon meter.

\begin{figure}
  \includegraphics[width=1\textwidth]{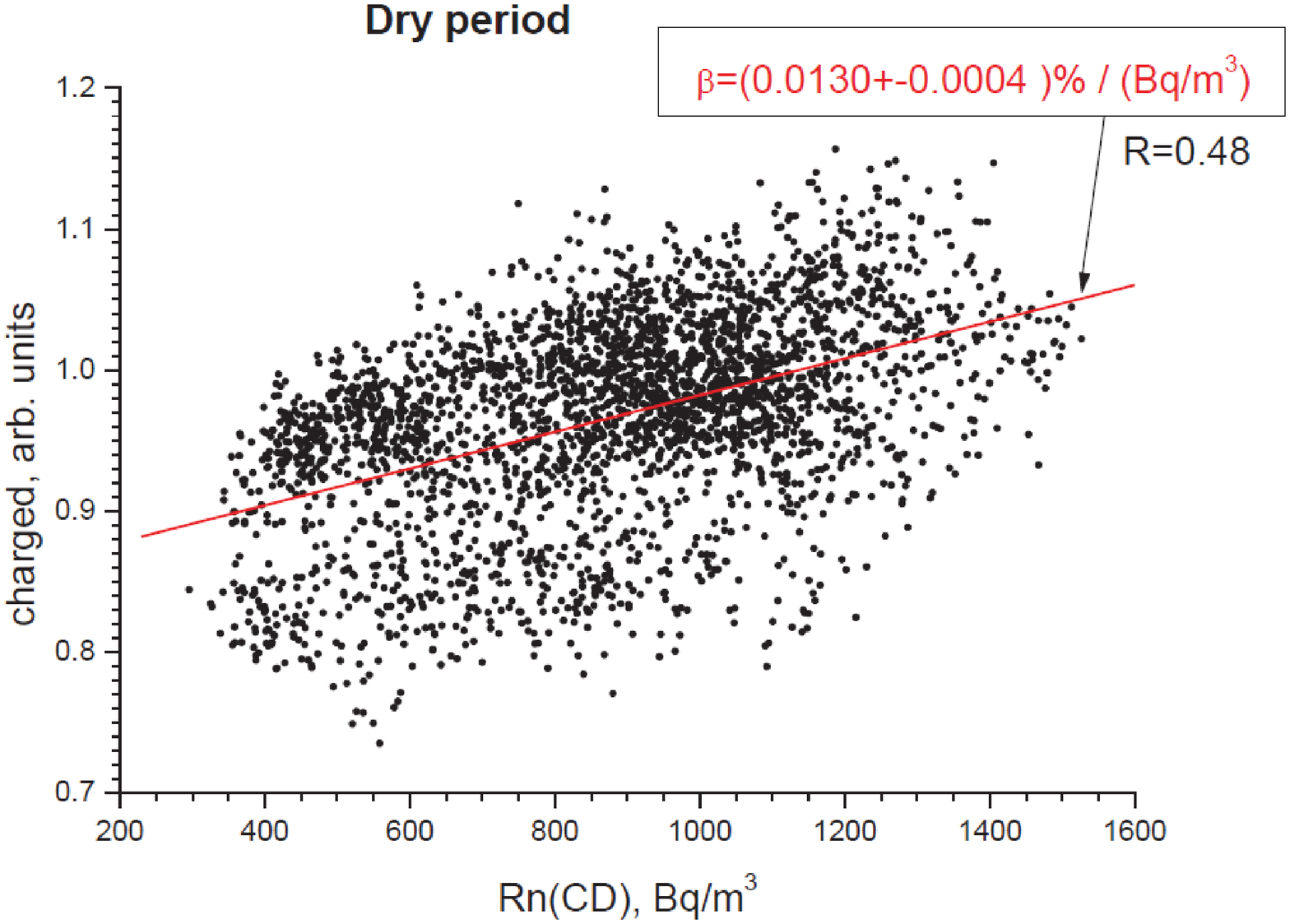}
  \includegraphics[width=1\textwidth]{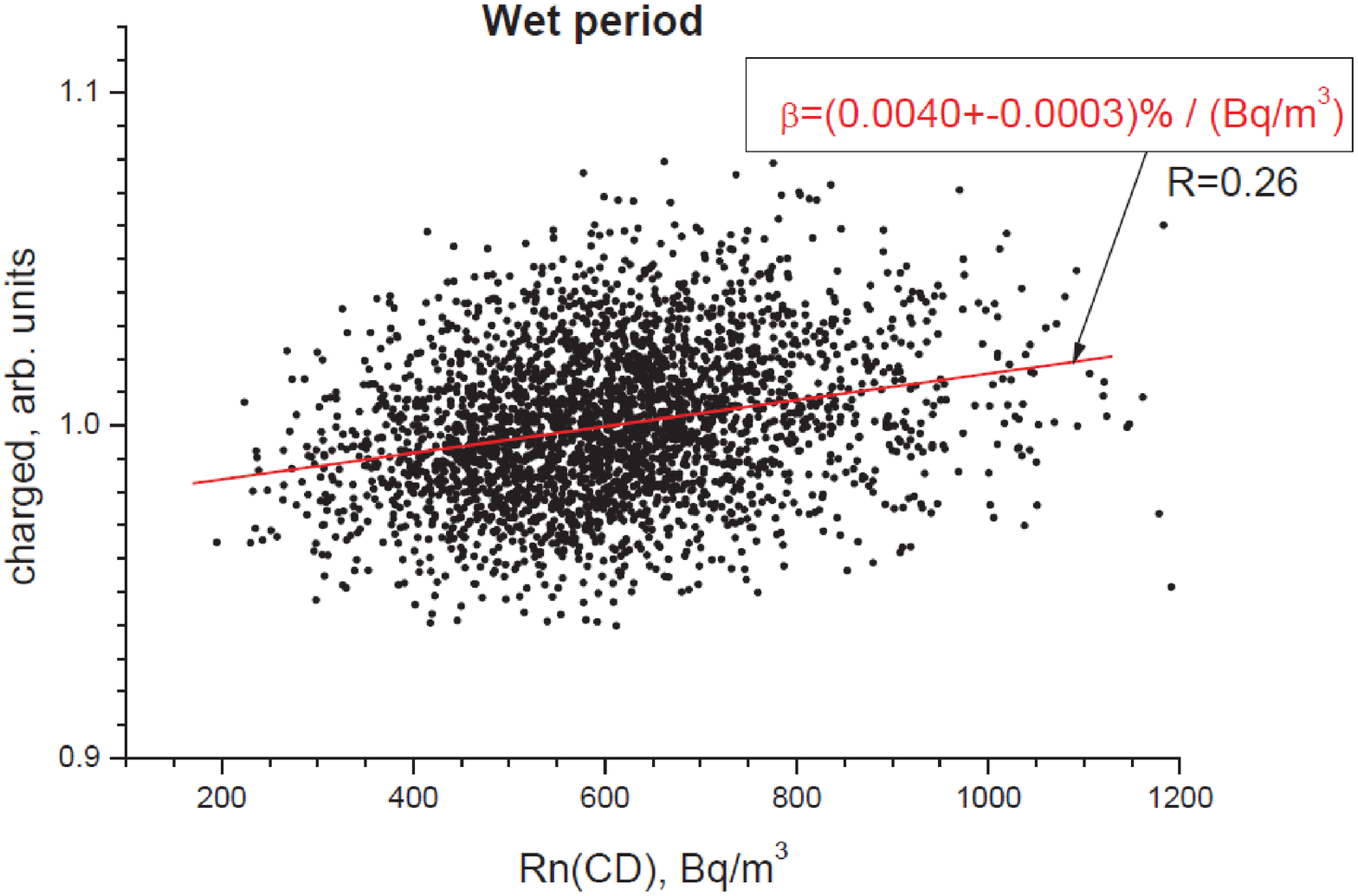}
  \caption{Correlation between radon air concentration and the "charged" channel in the driest season (February-March, upper) and in the wettest season (July-August, lower). The slope of the linear fit $\beta$ and the correlation coefficient R are shown.}
  \label{fig2}
 \end{figure}

In addition, we monitored radon concentration with 2 standard radon meters (Lucas cells, scintillation cells coated with zinc sulfide activated with silver, produced by MI.AM srl in Italy) located in the center of the hall and in its north edge, which stored the data every 30 min. These data along with locally measured meteorological data  were used to calibrate the en-detector "charged" channel to radon air volume activity. After 1.3 year of combined running, for the period from August 30, 2013 till December 3, 2014, we obtained the correlation between radon concentration in the hall and the "charged" channel, shown in Fig. \ref{fig2}. As the correlation depends on the air humidity we plotted separately the most dry (February-March) and the most wet (July-August) seasons.

\section{Results}

\subsection{Results on seasonal variations}

\begin{figure}
  \includegraphics[width=0.6\textwidth]{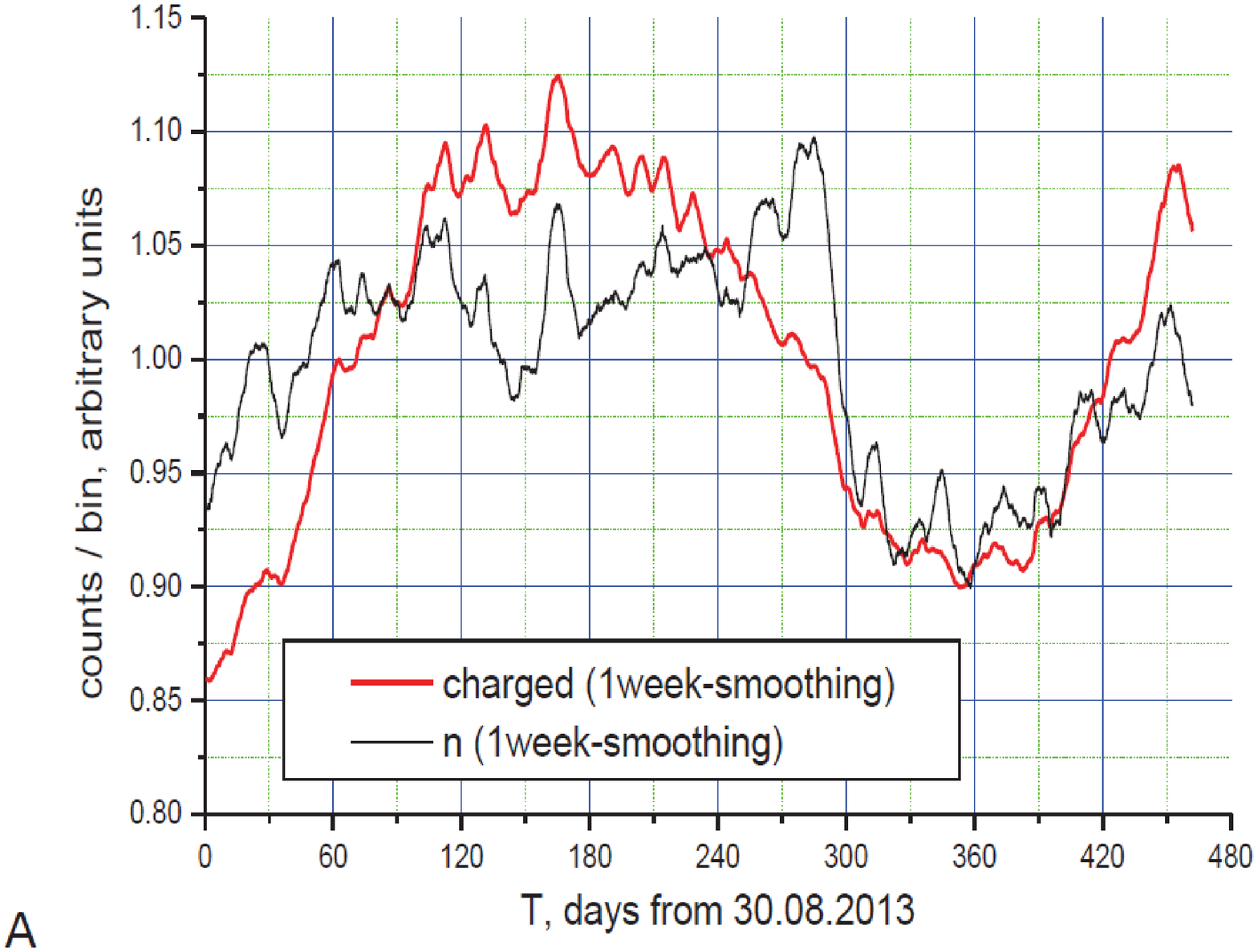}
  \includegraphics[width=0.6\textwidth]{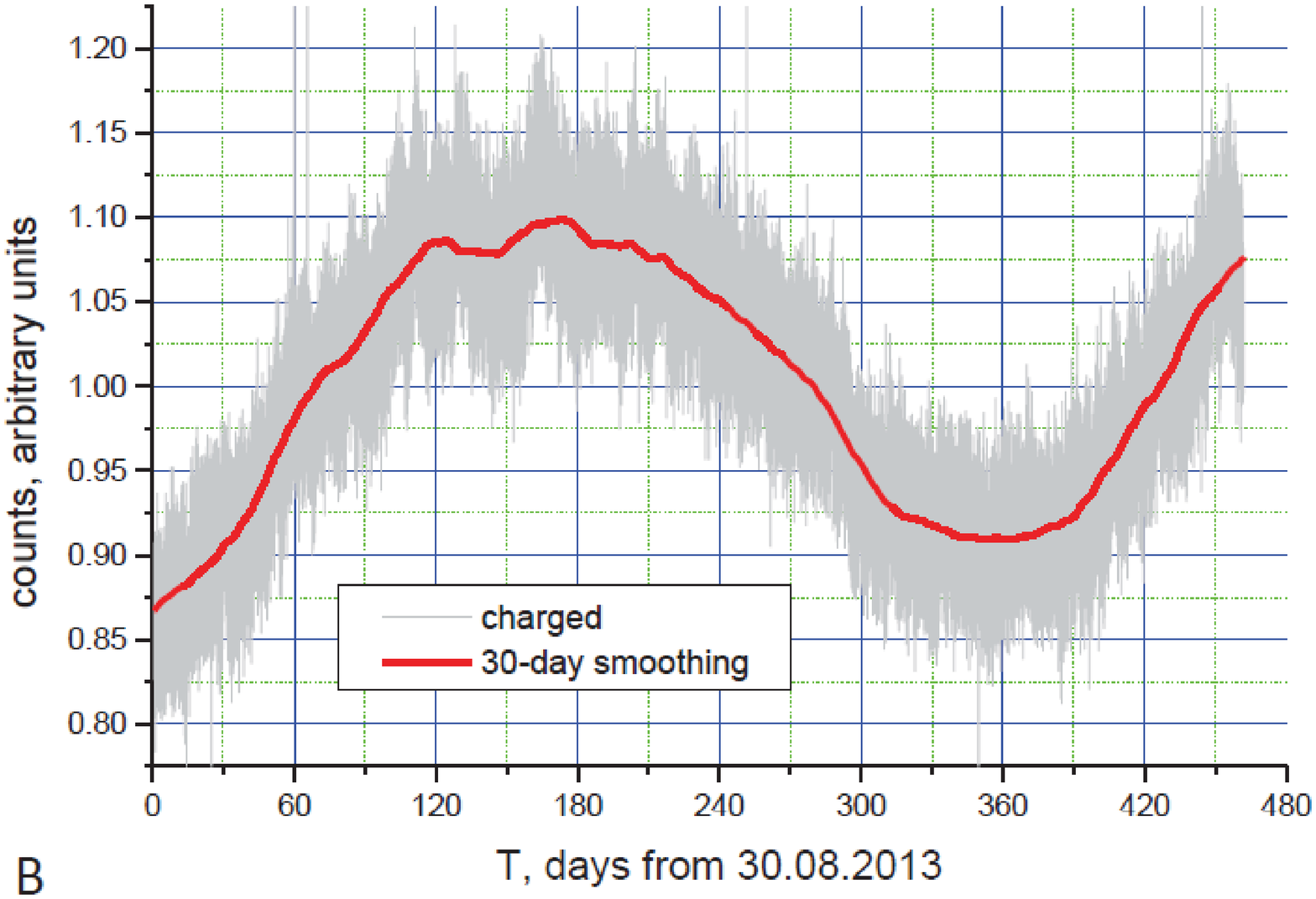}
  \includegraphics[width=0.6\textwidth]{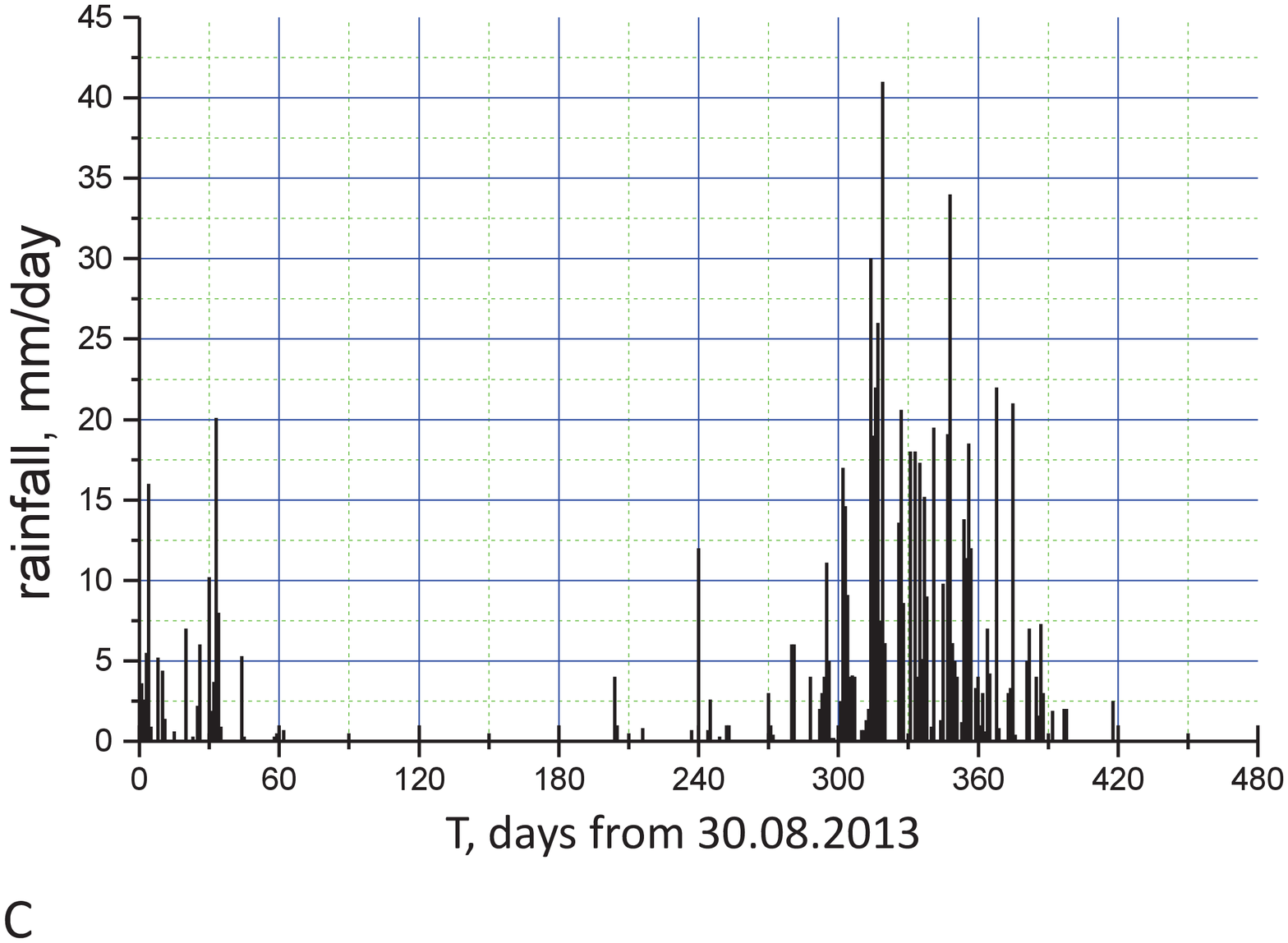}
  \includegraphics[width=0.6\textwidth]{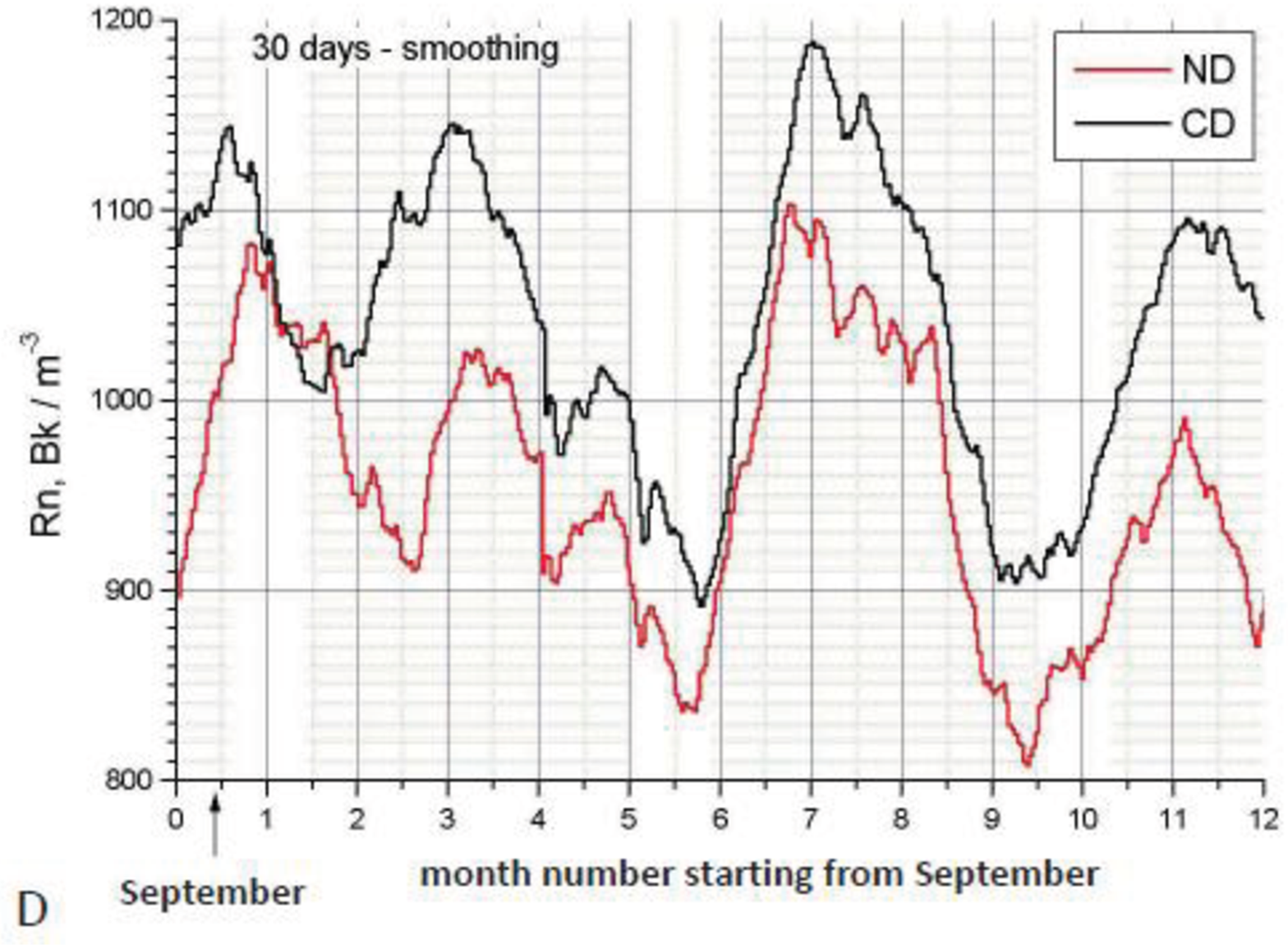}
  \caption{Seasonal effects for different parameters. Panel A- "neutrons" and "charged" channels with 1-week smoothing; panel B- "charged" channel with and without 1-month smoothing; panel C - rainfall in mm/day (taken from http://rp5.ru); panel D - air radon concentration measured by radon meters (Lukas cells) in two points of the ARGO-YBJ hall (ND - north and CD - central detectors) from Sep. till Aug. and averaged over 2 years.}
  \label{fig3}
 \end{figure}

Seasonal variations of the en-detector counting rate are shown in Fig. \ref{fig3} for pulses selected  as "neutrons" (black line) and as "charged" (red line) with 1-week smoothing (panel A). For the same period, data for the "charged" channel are shown in the panel B of Fig. \ref{fig3} without any correction as 5-minutes series (light gray) and as 30-day smoothing red line. Seasonal variations are evident in both channels.

The variations in the "charged" channel with 1-week (Fig. \ref{fig3}A) and particularly 1-month smoothing (Fig. \ref{fig3}B) are almost sinusoidal while those for "neutrons" are affected by air pressure, temperature and presence of precipitations, since water is a good neutron moderator and absorber. It can be seen that the fall of the neutron flux intensity (Fig. \ref{fig3}A) coincides with the beginning of the rain period in summer (Fig. \ref{fig3}C). This coincidence occurs for both types of signals while the maxima are not coincident but both occur during the dry season. Using the regression coefficient  from Fig. \ref{fig2}, the 10 \% seasonal wave of air radon measured through our "charged" channel corresponds to an amplitude of $\sim$ 770 Bq/m$^3$ in the dry season. Unfortunately we had only 1 dry and 1.5 wet periods at the date of the paper writing. That is probably why even at dry period the correlation coefficient is not high but difference between dry and wet periods is undoubtedly visible.

The radon concentration in air at high altitude is very high due to high natural radioactivity of surrounding rocks, high level of soil water containing a lot of Rn and activation of upper soil levels with cosmic rays hadrons increasing exponentially with altitude (ALEKSEENKO et al. 2011). The radon concentration in air measured through the Lucas cells (Fig. \ref{fig3}D) is affected by many random parameters as wind, ventilation, etc. and do not show a clear seasonal effect. On the contrary, underground radon measurements are more stable and results shown in (CIGOLINI 2009; ZMAZEK et al. 2003; FIRSTOV \& RUDAKOV 2015) confirm the existence of the seasonal effect on underground radon. Note that the amplitude of a seasonal wave in soil radon measured at Stromboli (CIGOLINI et al. 2009) was found to be of $\sim$ 2000 Bq/m$^3$ (or $\sim$ 50 \%) with a maximum coincident with the driest season.

The smoothing behavior of our "charged" data could be explained by the recording difference between Lucas cells and en-detectors: as mentioned above the "charged" channel is not sensitive directly to radon $\alpha$ decays but to cascades of $\gamma$s from $\beta$-decays of radon daughters (mostly $^{214}$Bi and $^{214}$Pb) close to the detector. Heavy metals like these are highly ionized and can actively attach to charged aerosols or dust that can be moved by wind to long distances. Even inside the ARGO-YBJ they can be deposited on the detector surface, made of plastic and therefore able to attract electrostatically the dust. Clearly, this process depends strongly on weather conditions like wind, rainfalls and air humidity. During the dry season this pollutants transport is greater while it is smaller during the rainy season. Note that the radon meter has an air filter on its inlet pipe making it not sensitive to dust and aerosols. This justifies the difference found between the "charged" and Lucas cells data.

The results show that the concentrations of underground radon and of thermal neutrons in equilibrium are higher during the dry period and that the difference between dry and wet periods could be of about 20 \% in radon-due thermal neutrons, while direct radon measurements in air gave between 30 \% and 40 \% for the seasonal variations in soil (ZMAZEK et al. 2003). Our "charged" counts due to the decays of radon daughters carried by dust shows a smooth seasonal wave of 10 \% amplitude.  We should note here that these counts are produced not only by radon decays but also by other processes, such as small EASs, interactions of cosmic rays just above the detector and even PMTs noise. It is probably for this reason that we found an amplitude of the "charged" counts seasonal wave smaller than one obtained with direct radon measurements.

\subsection{Lunar month variations}

\begin{figure}
  \includegraphics[width=1\textwidth]{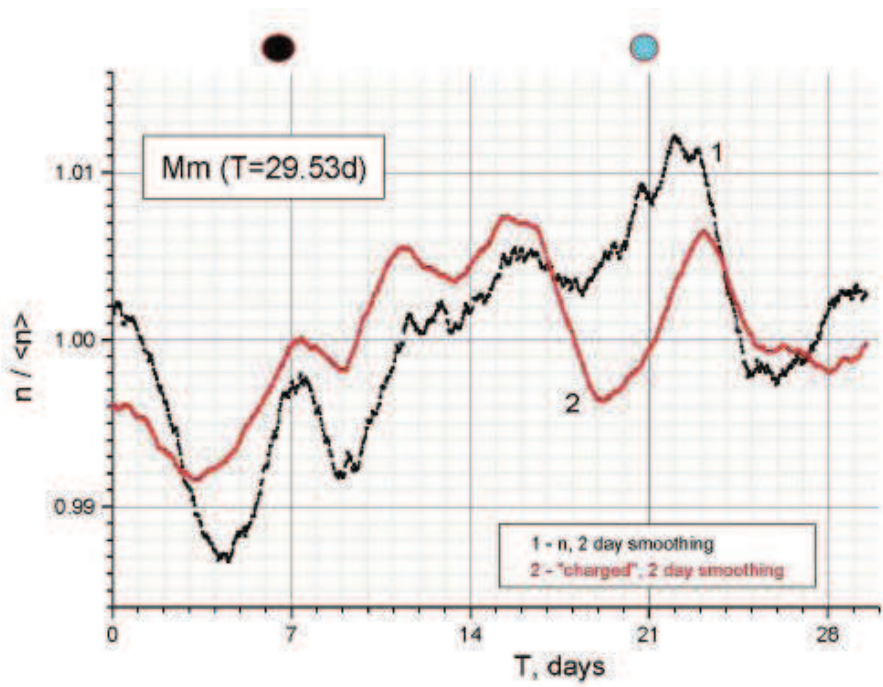}
  \caption{Superimposed epoch analysis applied to the neutron and "charged" data for the synodic moon month. Black circle: new moon; cyan circle: full moon.}
  \label{fig4}
 \end{figure}

It is believed that deformations of the Earth crust can induce radon exhalation from the ground, where radon is produced continuously due to decays of Uranium and Thorium radioactive chains, to the atmosphere. These changes can be triggered by geophysical processes as earthquakes, moon and solar tides and free Earth oscillations. The existence of moon tidal waves in radon concentration in soil and underground laboratories has been established many years ago (see for example (ALEKSEENKO et al. 2009; MAJERUS and LANTREMANGE 1999; GROVES-KIRKBY et al. 2004; RICHON et al. 2012). In this study we applied the superimposed epoch and the Fast Fourier Transformation (FFT) analysis to the neutron and "charged" time series. Superimposed epoch analysis for full time series has also been used for known lunar periods to show the shape of the corresponding wave with 2-days smoothing. Fig. \ref{fig4} shows the normalized result of the superimposed epoch analysis (i.e. sum of many overlapped periods) for the synodic moon month (Mm, period = 29.53 days). This picture displays rather interesting wave structures showing clearly the 1st harmonic with a maximum almost coincident with Full Moon and an amplitude of $\sim$ 1 \%, corresponding to $\sim$ 77 Bq/m$^3$ in dry season and higher harmonics as well.

\begin{figure}
  \includegraphics[width=1\textwidth]{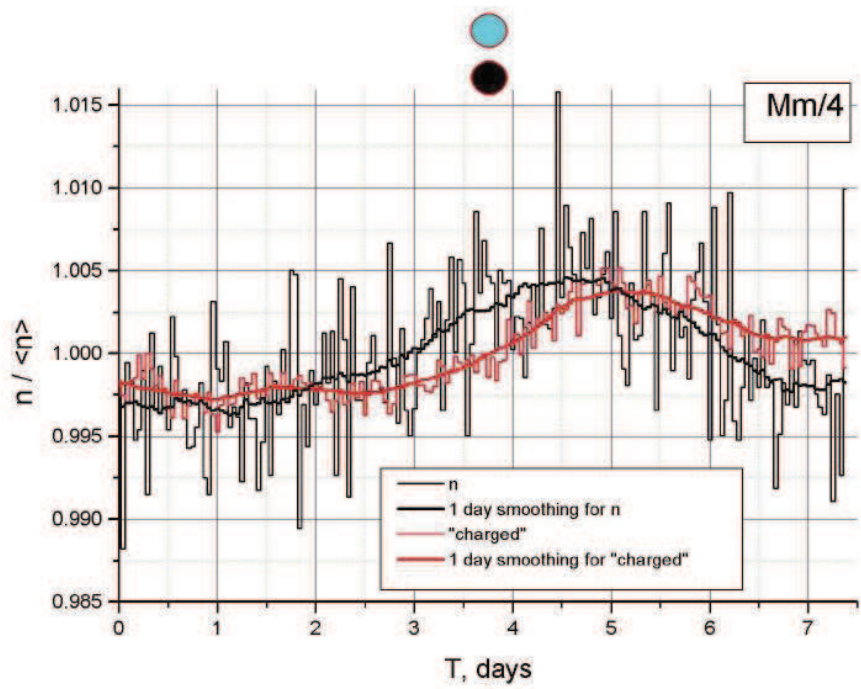}
  \caption{Superimposed epoch analysis applied to neutrons and "charged" data for one quarter of the synodic moon month.}
  \label{fig5}
 \end{figure}

To confirm the existence of the 4th harmonic we made the same analysis for 1/4 of moon period (Mm/4), corresponding to about 1 week using 1 hour data points. The result (Fig. \ref{fig5}) shows a clear maximum confirming the 4th harmonic existence of about 5 \% amplitude. This indicates that radon and neutron excesses occur not only in coincidence with moon syzygies but also with lunar quarters. This phenomenon could be due to the fact that during syzygies, Moon, Earth and Sun are aligned (both for New and Full Moon) and the tidal forces from Moon and Sun are added up. This generates the maxima just after the syzygies. On the other hand, halfway between two syzygies the forces from Moon and Sun are perpendicular to each other producing a deformation of the crust but in orthogonal directions. The amplitude of this variation is about a factor of 2 lower than the corresponding value for straight directions.

Similar to seasonal variations the neutron counts show a sharper variation compared to the "charged" channel with a general good agreement. This correlation between neutron and "charged" channels for the synodic moon month analysis is probably due to the fact that the delay of several hours (if any) is not visible on a scale of 30 days with a 2-days smoothing. Looking to Fig. \ref{fig5}  (7.4 days with 1-day smoothing) a delay of maxima with respect to syzygies of one day for neutrons and $\sim$ 1.5 days for "charged" is clearly visible. The amplitude for both is close to 0.5 \%.

\begin{figure}
  \includegraphics[width=1\textwidth]{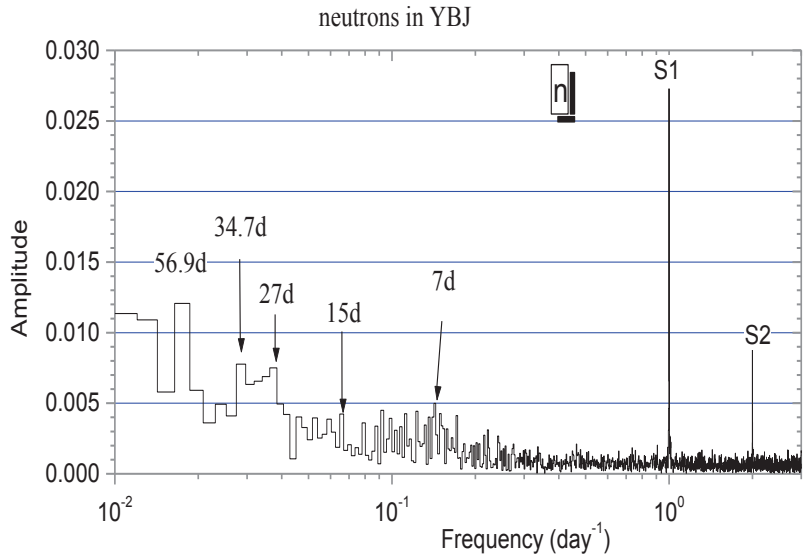}
  \includegraphics[width=1\textwidth]{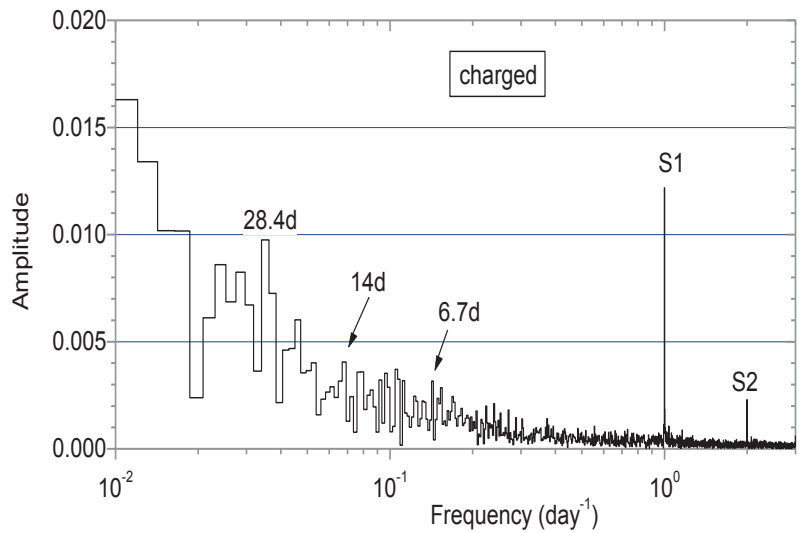}
  \caption{ Results of the FFT analysis of 5 minutes time series covering 480 days of data taking for neutrons (upper) and "charged" (lower).}
  \label{fig6}
 \end{figure}

The same neutron and "charged" data have been analyzed by means of the FFT analysis (Fig. \ref{fig6}) using full data set of 5 minutes time series. The plot shows the observed frequency amplitude vs frequency. For both channels the diurnal wave S1 (due to air temperature) and semi diurnal wave S2 are clearly visible, with an amplitude for neutrons $\sim$ 2 times greater than for "charged". Moon month periods and their harmonics are also seen clear in both graphs, with a greater neutron amplitude that confirms the results shown in Fig. \ref{fig4}. It should be noticed that there are several lunar month periods such as synodic, draconic, anomalistic, etc. with close periods in the range of 27.2 - 29.5 days, similar with solar period of $\sim$ 27 days, but longer data time series have to be added up to distinguish between them. Up to now we have only 15 months of data, not enough to separate them in a statistically significant way. Nevertheless, we certainly see in both distributions moon month periods in both distributions together with different peaks due to different behavior of neutrons and "charged" in Fig. \ref{fig3}-\ref{fig5}, with "charged" data smoother than neutron ones.
	
\begin{figure}
  \includegraphics[width=1\textwidth]{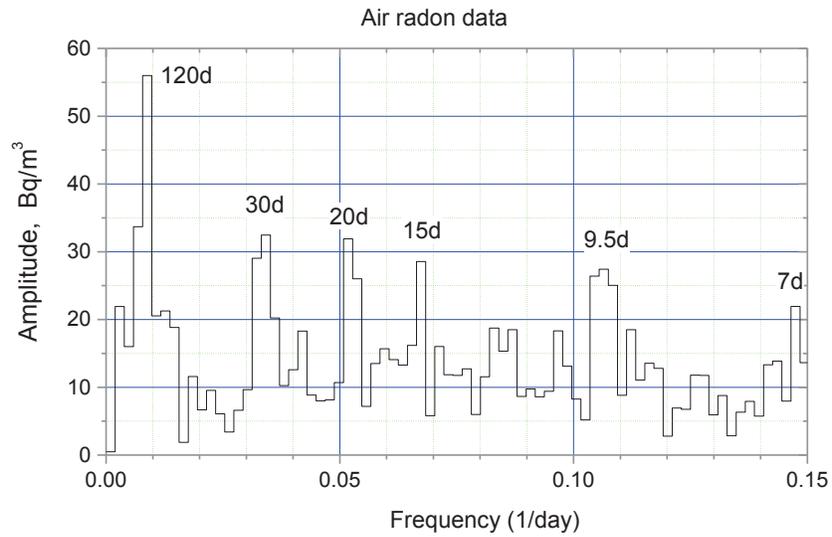}
  \includegraphics[width=1\textwidth]{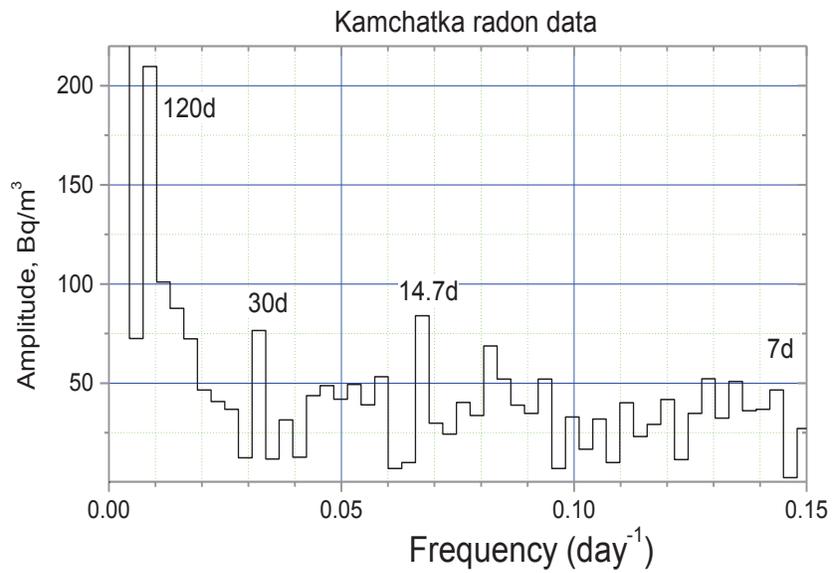}
  \caption{Result of FFT analysis for radon data in air (Tibet, upper) and in soil (Kamchatka, lower).}
  \label{fig7}
 \end{figure}

Fig. \ref{fig7} shows the results of the FFT analysis for data obtained with air radon meters in YangBaJing (upper panel) and in soil at Kamchatka (FIRSTOV 2015)(lower panel). Both results are obtained for $\sim$ 450 days period of data taking. A qualitative agreement between all the data shown in Fig. \ref{fig6} and \ref{fig7} can be seen, with moon month periodicities from 27 to 30 days visible in all graphs. This confirms at least partially the radon origin of the natural thermal neutron flux variations. Our "charged" and neutron data are less sensitive to some periods well seen in the air radon data, such as 20 days and 9.5 days, not seen also in soil radon data at Kamchatka. On the other hand, the data (FIRSTOV 2015)show the existence of a $\sim$ 120 days period confirming the authenticity of this period seen also in our air radon data (Fig. \ref{fig3}D).

\section{Discussion and Conclusions}

New results are obtained using a new nuclear physics method based on continuous measurement of: 1) $^{222}$Rn concentration in air; 2) $^{222}$Rn daughters in dust (aerosols) and 3) $^{222}$Rn flux density in soil through the detection of thermal neutrons produced in ($\alpha$, n)-reactions underground by $\alpha$s originated from the radon decay chain. The existence of periodical variations both seasonal and related to moon month are confirmed for the first time at high altitude. In particular, for synodic moon month the 4th harmonic with amplitude of about 0.5 \% and maximum with a delay of about 1 day after syzygies is found. Notice that the existence of these periodicities at low altitude was claimed in our previous paper (ALEKSEENKO et al. 2009), but it is the first indication of the existence of a $\sim$ 7.4 day period in the radon flux density both in soil and in air due to lunar month tidal waves. Moreover, the existence of the seasonal effect for radon and thermal neutrons in anti-correlation with the rain season is now proved. Taking into account that our detector is running at the surface and estimated fraction of radon originated neutrons is about 15 \% (the rest 85 \% from cosmic rays), then observed amplitudes could be recalculated to soil radon amplitudes  through a factor of $\sim$ 7. Therefore, the 4th harmonic of moon month wave in soil radon is expected to be $\sim$ 3.5 \% and that for seasonal wave is about 52 \% in agreement with that obtained in direct soil radon measurements (CIGOLINI et al. 2009).

According to (YAKOVLEVA 2003), the radon flux density underground is much more sensitive than concentration to Earth crust distortions. Our new method developed to study neutrons originated from radon is sensitive to the number of nuclides per unit of volume and time from the radon chain decays in soil, concrete, or other similar materials close to the detector. This means that we can monitor the flux density of radon underground since radon is the main carrier of natural radioactivity. The advantages of this thermal neutron method are: long term stability, insensitivity to wind and ventilation, sensitivity to underground radon up to several meters without drilling any hole and sensitivity to tidal waves. This gives us a hope to develop a method for earthquakes prediction provided by a global net of en-detectors (ALEKSEENKO et al. 2013; ALEKSEENKO et al. 2015). (sometimes we do see the en-detectors response to nearby earthquakes but this is a subject of our future papers.)

\section{acknowledgments}
This work was supported in Russia by RFBR (grants 14-02-00996 and 13-02-00574), RAS Presidium Program "Fundamental properties of matter and astrophysics", and in China by NSFC (No.10975046, No.11375052). We also acknowledge the support of the ARGO-YBJ collaboration.

ALEKSEENKO, V. V., ARNEODO, F. et. al. (2015). Decrease of Atmospheric Neutron Counts Observed during Thunderstorms. Physical Review Letter, 114, 125003. \\ \\
ALEKSEENKO, V. V., ARNEODO, F., BRUNO, G. et al. (2013). Sporadic Variations of Thermal Neutron Background Measured by a Global Net of the En-detectors. the 33rd International Cosmic Ray Conference  proceeding, Rio De Janeiro, ID 568. \\ \\
ALEKSEENKO, V. V., GAVRILYUK, Yu. M., GROMUSHKIN, D. M. et al. (2009). Correlation of Variations in the Thermal Neutron Flux from the Earth's Crust with the Moon's Phases and with Seismic Activity. Izvestiya, Physics of the Solid Earth, Vol. 45, No. 8, pp. 709-718. \\ \\
ALEKSEENKO, V. V., GAVRLYUK, Yu. M., KUZMINOV V. V. and STENKIN, Yu. V. (2010). Tidal Effect in the Radon-due Neutron Flux from the Earth's Crust. Journal of Physics: Conference Series. 203, 012045 \\ \\
ALEKSEENKO, V. V., GROMUSHKIN, D. M. and STENKIN, Yu. V. (2011). Comparative Measurements of Thermal Neutron Fluxes at Ground Level at the Baksan Neutrino Observatory and LGNS Laboratory. Bulletin of the Russian Academy of Sciences. Physics, Vol. 75, No. 6, pp. 857-859. \\ \\
BARTOLI, B. et al. (2016). Detection of Thermal Neutrons with the PRISMA-YBJ Array in Extensive Air Showers Selected by the ARGO-YBJ Experiment. Accepted by Astroparticle Physics, DOI: 10.1016/j.astropartphys.2016.04.007; arXiv:1512.01326 [astro-ph.IM].\\ \\
CIGOLINI, C., POGGI, P., RIPEPE, M. et al. (2009). Radon Surveys and Real-time Monitoring at Stromboli Volcano: Influence of Soil Temperature, Atmospheric Pressure and Tidal Forces on $^{222}$Rn Degassing. Journal of Volcanology and Geothermal Research, 184 381-388. \\ \\
D'ETTORRE BENEDETTO, P. on behalf of the ARGO-YBJ collaboration (2011). Highlights from the ARGO-YBJ Experiment, The 32nd International Cosmic Ray Conference proceeding, Beijing, v12, 93-106. \\ \\
FIRSTOV, P. P. (2015). Kamchatka Branch of Geophysical Servey of Russian Academy of Sciences, private communications. \\ \\
FIRSTOV, P. P. and RUDAKOV, V. P. (2003). Results of Recording of Subsurface Radon in 1997-2000 at the Petropavlovsk Kamchatski Geodynamic Research Area. Vulkanologiya i Seismologiya 1: 26-41. \\ \\
GROVES-KIRKBY, C. J., DENMAN, A. R. et al. (2004). Periodicity in Domestic Radon Time Series - Evidence for Earth Tides. Proc. IRPA'11, Madrid. \\ \\
PRISMA-YBJ collaboration 2013. Coincident air shower events between ARGO-YBJ and PRISMA-YBJ, \textit{The 33rd ICRC proceeding}, \textit{Rio de Janeiro}, ID 606.
MAJERUS, KIES A.J., and DE LANTREMANGE, N. D'OREYE (1999). Underground Radon Gas Concentrations Related to Earth Tides, Nuovo Cimento Soc. Ital. Fis. C, 22, 287-293. \\ \\
RICHON, P., MOREAU, L. et al. (2012). Evidence of both M2 and O1 Earth Tide Waves in Radon-222 Air Concentration Measured in a Subglacial Laboratory, Journal of Geophysical Research, VOL. 117, B12404. \\ \\
STENKIN, Yu. V. (2010) Large Scintillator Detector for Thermal Neutron Recording. Nuclear Track Detectors: Design, Methods and Applications ISBN: 978-1-60876-826-4, Editor: Maksim Sidorov and Oleg Ivanov, Nova Science Publishers, Inc., Chapter 10, p. 253-256. \\ \\
YAKOVLEVA, V. S. (2003). The Radon Flux Density from the Earth¡¯s Surface as an Indicator of a Seismic Activity. Proceedings of ICGG7, Copernicus GmbH, 28-30. \\ \\
ZMAZEK, B.,  TODOROVSKI, L. et al. (2003). Application of Decision Trees to the Analysis of Soil Radon Data for Earthquake Prediction. Applied Radiation and Isotopes, 58, 697-706. \\ \\

\end{document}